# Emergence of Seismic Metamaterials: Current State and Future Perspectives

Stéphane Brûlé[1], Stefan Enoch[1] and Sébastien Guenneau[1]

[1] *Aix Marseille Univ, CNRS, Centrale Marseille, Institut Fresnel, Marseille, France*
52 Avenue Escadrille Normandie Niemen, 13013 Marseille
e-mail address: sebastien.guenneau@fresnel.fr

Following the advent of electromagnetic metamaterials at the turn of the century, researchers working in other areas of wave physics have translated concepts of electromagnetic metamaterials to acoustics, elastodynamics, as well as to heat, mass and light diffusion processes. In elastodynamics, seismic metamaterials have emerged in the last decade for soft soils structured at the meter scale, and have been tested thanks to full-scale experiments on holey soils five years ago. Born in the soil, seismic metamaterials grow simultaneously on the field of tuned-resonators buried in the soil, around building's foundations or near the soil-structure's interface, and on the field of above-surface resonators. In this perspective article, we quickly recall some research advances made in all these types of seismic metamaterials and we further dress an inventory of which material parameters can be achieved and which cannot, notably from the effective medium theory perspective. We finally envision perspectives on future developments of large scale auxetic metamaterials for building's foundations, forests of trees for seismic protection and metamaterial-like transformed urbanism at the city scale.

## I. INTRODUCTION

In Civil Engineering, the high density of deep foundation or ground reinforcement techniques for buildings in urban area, leads Physicists to believe in a significant interaction of these buried structures (FIG. 1) with a certain component of the seismic signal. In the past, a few authors ([**1**] and [**2**]) obtained significant results with vibration screening in the soil itself for a local source such as industrial vibratory machines located on concrete slab for example.

Léon Brillouin (1946) reminded that « *All waves behave in a similar way, whether they are longitudinal or transverse, elastic or electric* » [**3**]. However, to illustrate the interaction of seismic wave with structured soils, researchers had to bring specific theoretical and original experimental approaches in particular because of the complexity of the wave propagation in the Earth's surface layers ([**4**] and [**5**]). In this article we illustrate how the wave interaction concept had been extended this last decade to seismic waves generated by earthquakes [**6**].

## II. OVERVIEW OF PHOTONIC, PHONONIC, PLATONIC CRYSTALS AND LINKS TO SEISMIC METAMATERIALS

In 1987, the groups of E. Yablonovitch and S. John reported the discovery of stop band structures for light ([**7**] and [**8**]). Photonic crystals (PCs) have, since then, found numerous applications ranging from nearly perfect mirrors for incident waves whose frequencies are in stop bands of the PCs, to high-q cavities for PCs with structural defects [**9**].

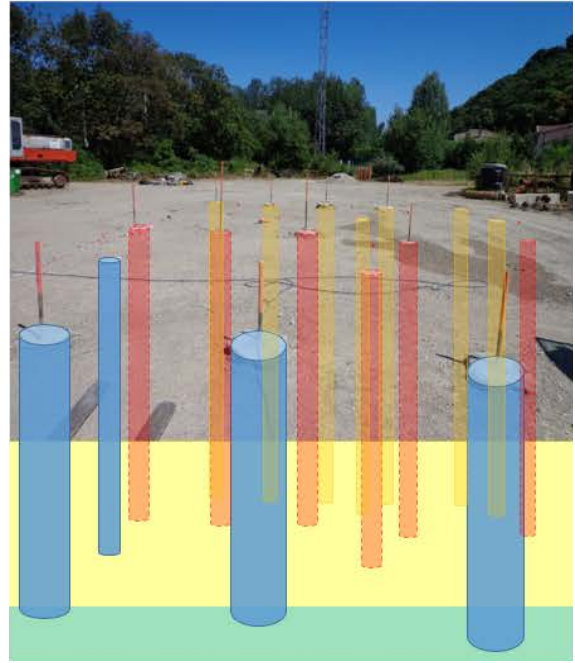

FIG. 1. Principle of 2D artificial and structured soil made of a mesh of cement columns clamped in a substratum (green layer).

The occurrence of stop bands in PCs also leads to anomalous dispersion whereby dispersion curves can have a negative or vanishing group velocity. Dynamic artificial anisotropy, also known as all-angle-negative-refraction

([10] to [13]), allows for focusing effects through a PC, as envisioned in the 1968 paper of V. Veselago [14]. With the advent of electromagnetic metamaterials ([15] and [16]), J. Pendry pointed out that the image through the V. Veselago lens can be deeply subwavelength [17], and exciting effects such as simultaneously negative phase and group velocity of light [18], invisibility cloaks [19] and tailored radiation phase pattern in epsilon near zero metamaterials were demonstrated [20] and [21]. One of the attractions of platonic crystals, which are the elastic plate analogue of photonic and phononic crystals, is that much of their physics can be translated into platonics.

There are mathematical subtleties in the analysis, and numerics, of the scattering of flexural waves [22] owing to the fourth-order derivatives in the plate equations, versus the usual second-order derivatives for the wave equation of optics, involved in the governing equations; even waves within a perfect plate have differences from those of the wave equation as they are not dispersionless. Nonetheless, drawing parallels between platonics and photonics helps to achieve similar effects to those observed in electromagnetic metamaterials, such as the time dependent subwavelength resolution through a platonic flat lens [23].

In parallel, research papers in phononic crystals provided numerical and experimental evidence of filtering [24] and focusing properties [25] of acoustic waves.

Localized resonant structures for elastic waves propagating within three-dimensional cubic arrays of thin coated spheres [26] and fluid filled Helmholtz resonators [27] paved the way towards acoustic analogues of electromagnetic metamaterials ([28] and [29]), including elastic cloaks ([30] to [32]). The control of elastic wave trajectories in thin plates was reported numerically [33] and experimentally in 2012 [4] and 2014 [5] for surface seismic waves in civil engineering applications. In fact, Rayleigh waves are generated by anthropic sources such as an explosion or a tool impact or vibration (sledge-hammer, pile driving operations, vibrating machine footing, dynamic compaction, etc.).

In 1968, R.D. Woods [1] created *in situ* tests with a 200 to 350 Hz source to show the effectiveness of isolating circular or linear empty trenches, with the same geometry, these results were compared in 1988 with numerical modeling studies provided by P.K. Banerjee [2].

The main thrust of this article is to point out the possibility to create seismic metamaterials not only for high frequency anthropic sources but for the earthquakes' frequency range i.e. 0.1 to 12 Hz.

## III. WAVE PROPAGATION IN EARTH SUPERFICIAL LAYERS

While the mathematical laws describing seismic waves are those of elastodynamics (Navier's equation), there are complexities and scalability issues that make the metamaterial transitions into the "solid Earth" realm a very challenging problem. Firstly, seismic waves have a wavelength of the order of tens if not hundred of metres at the frequencies relevant in civil engineering. Secondly the geological and seismological characteristics of the Earth's superficial layers of the crust drastically increase the complexity of the wavefield.

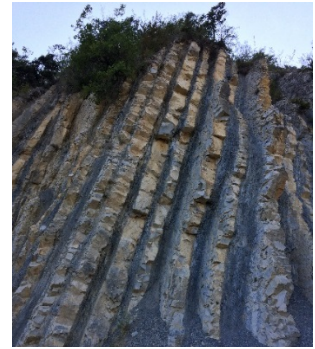

FIG. 2. Outcrop illustrating the alternation of clay (soft) –calcareous (rigid) layers reminiscent of 1D metamaterial. Serres, Drôme, France (courtesy of S.Brûlé).

Indeed, the incoming seismic signal could generate several types of surface waves and guided modes especially when waves are trapped in sedimentary basins (so-called basin resonance and seismic site effects). Finally, given the strong motion that usually accompanies seismic events, it is yet unknown the role of the soil-metamaterial interactions and the effect played by non-linearity's and attenuation. We consider the case of constructions sufficiently far from the zones of tectonic faults to consider that the conditions of the elasticity are applicable.

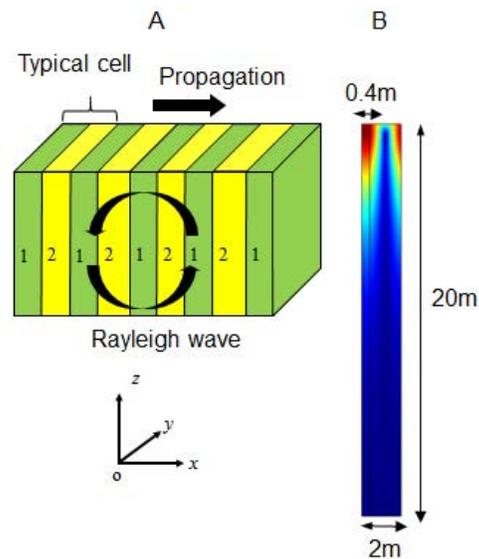

FIG. 3. (A) Schematic diagram of a vertical-layered and periodic geological structure (1D); (B) Simulation of a Rayleigh Bloch wave propagating within a soil (thick layer)-concrete (thin layer) periodic medium. Color scale ranges from dark blue (vanishing amplitude of displacement field) to red (maximum amplitude). Note the contrast in material parameters makes this a locally resonant structrure.

The surface wave is the main composition of wave energy in ground vibration, so researchers recently suggest the use of layered periodic structure to mitigate ground vibrations propagating in the soil [34]. As depicted in FIG. 3 (A) the structured soil is layered with two different engineering materials. Let us consider an alternation of layers with and concrete and soil parameters, a Rayleigh Bloch wave emerges as shown in the numerical simulation in FIG. 3 (B).

Due to its periodicity, the typical cell (see Fig. 3(A)) is commonly used to analyze the dynamic characteristics of infinitely extended periodic structures. Both layers in the typical cell are assumed to be homogeneous, isotropic, and perfectly bonded at the interface. Ignoring the body force, the harmonic motion of a 2D plane strain typical cell (Eq. 1) can be drawn as follows [35]:

$$\frac{1}{\rho(r)}\left[\frac{\partial}{\partial x_i}\left(\lambda(r)\frac{\partial u_j}{\partial x_j}\right) + \frac{\partial}{\partial x_j}\left(\mu(r)\left(\frac{\partial u_i}{\partial x_j} + \frac{\partial u_j}{\partial x_i}\right)\right)\right] = -\omega^2 u_i^2 \ (i,j = x,z)$$

(1)

where u, ρ, λ, and μ are displacement, density, and two Lame's constants of the mediums, respectively; r is the displacement vector and ω is the circular frequency. These periodic barrier made of open trench, polyfoam in-filled trench, rubber in-filled trench or concrete panel [34] are effective for signal of few tens of hertz (railways vibrations, etc.). Further note that the leftmost and rightmost vertical sides of the periodic cell in Fig. 3(B) are supplied with Floquet-Bloch boundary conditions that induce a phase shift like $u_i(x+l,z) = u_i(x,z)e^{ikl}$ where l is the array pitch and k the Bloch wavenumber that lies in the irreducible Brillouin zone $[0, \pi/l]$. Further note that Love waves, which are polarized in the out-of-plane direction y, cannot propagate in such a medium as there is no propagating sublayer below the surface.

## IV. SCIENTIFIC BREAKTROUGH IN EARTHQUAKE ENGINEERING AND SOIL-STRUCTURE INTERACTION

Most of the vibration energy affecting nearby structures is carried by Rayleigh surface waves. Earthquake Engineering is concerned with the horizontal component of bulk and surface waves [4].

The response of a structure to earthquake shaking is affected by interactions between three linked systems: the structure, the foundation, and the soil underlying and surrounding the foundation. Soil-structure interaction analysis evaluates the collective response of these systems to a specified ground motion. The terms Soil-Structure Interaction (SSI) and Soil-Foundation-Structure Interaction (SFSI) are both used to describe this effect ([36]).

The term free-field refers to motions that are not affected by structural vibrations or the scattering of waves at, and around, the foundation. SSI effects are absent for the theoretical condition of a rigid foundation supported on rigid soil. However, in the case of structured soils and, among them, seismic metamaterials, we are specifically looking to interact with high concentration of foundations in the soil (piles, retaining walls, inclusions, etc.).

Earthquake Engineers can act in different ways for the design of structures.

In most of the cases, they can directly use the input data from the local regulation which is the result of a deterministic or probabilistic analysis of the seismic hazard.

From an elastic approach of the structure, engineers usually introduce frame ductility, seismic insulators or soil-structure interaction, to reduce the pseudo-static loads applied or to improve the mechanical response of the structure.

Most of the time in earthquake engineering, the dynamic response of a building is formulated as an elastic response spectrum S(T, ξ), in acceleration $S_a$, velocity $S_v$ or horizontal displacement $S_d$ of a system (mass m, stiffness k) simplified as a single degree of freedom oscillator (SDOFO) under seismic acceleration of the soil $\ddot{x}_g(t)$.

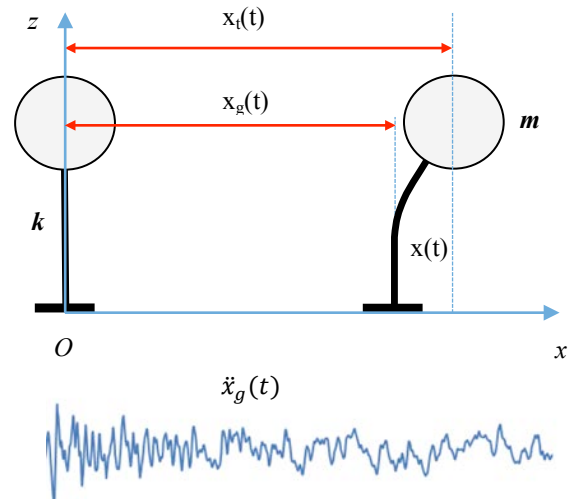

FIG. 4. Principle of usual single degree of freedom oscillator (SDOFO) model for a building (From [**37**]).

From each seismogram, earthquake engineers point out the maximum value of acceleration, velocity or displacement, for a relative displacement x(t) in time domain, for given damping ratio $\xi_0$ and a fundamental period T of the structure (FIG. 4) and natural pulsation $\omega_0$. Duhamel's integral is a way of calculating the response of linear structures x(t) which is the steady state solution, to arbitrary time-varying external perturbation $\ddot{x}_g(t)$ (Eq. 2).

$$x(t) = \frac{1}{\omega_d} \int_{\tau=0}^{\tau=t} \ddot{x}_g(\tau) e^{-\xi_0 \omega_0 (t-\tau)} \sin(\omega_d (t-\tau)) \, d\tau \quad (2)$$

Here, $x_t(t)$ is the overall displacement for a visco-elastic system and $x_g(t)$ is the displacement of the soil deduced from the records of surface seismograph. $\omega_d$ is the damped pulsation of the system.

Except for a few cases with full numerical modeling, the complete study of both the structure and deep foundations (piles) is complex and usually the problem is split into two: kinematic and inertial interactions (FIG. 5).

Kinematic interaction results from the presence of stiff foundation elements on or in soil, which causes motions at the foundation to deviate from free-field motions. Inertial interaction refers to displacements and rotations at the foundation level of a structure that result from inertia-driven forces such as base shear and moment.

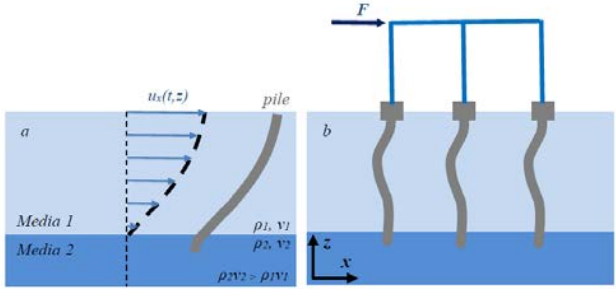

FIG. 5. Earthquake Engineering for deep foundations. Kinematic interaction (left) and inertial effect (right). Inspired from [**36**].

The first novelty of this last decade is the development of the interaction of structured soil with the seismic signal propagating in superficial Earth's layers [**38**] and then an active action on the kinematic effect described above. The second advance is the experimental modal analysis of composite systems made of soil, piles and structures by means of numerical analysis and the confirmation of these predictions by full-scale experiments held in situ.

## V. DIFFERENT TYPES OF SEISMIC METAMATERIALS

We could identify three types of seismic metamaterials after a decade of research.

**Seismic Soil-Metamaterials**

The first type includes structured soils made of cylindrical voids ([**4**], [**5**] and [**40**]) or rigid inclusions ([**41**] and [**42**]), including seismic metamaterials. We can call this historical first group, "Seismic Soil-Metamaterials" or SSM (FIG. 6).

These full-scale experiments with cylindrical holes allowed the identification of the Bragg's effect and the distribution of energy inside the grid (FIG. 6), which can be interpreted as the consequence of an effective negative refraction index. Such a flat lens reminiscent of what Veselago and Pendry envisioned for light.

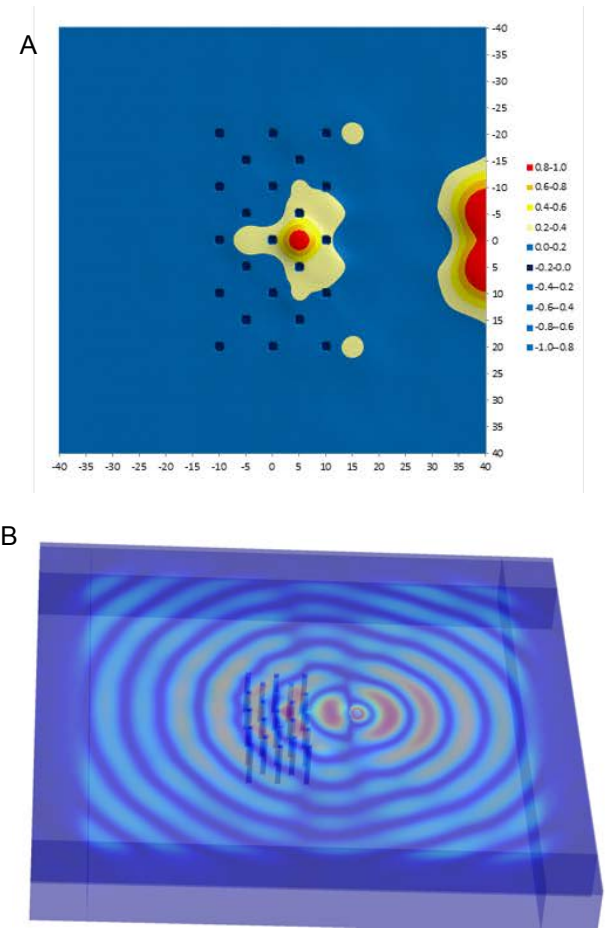

FIG. 6. (A) Snapshot illustrating the recorded energy distribution in a 23 holes structured soil after an impact at the Earth's surface [**5**]. Source is located at (x=-40, y=0) with a mean frequency content around 8 Hz. (B) Simulation (comsol) of an elastic wave generated by a vertical point

force in front of 23 stress-free inclusions (depth 50m, diameter 2m, center-to-center spacing 7m) in a thick plate (thickness 50m) with soil parameter (Young modulus E=0.153 GPa, Poisson ratio ν=0.3, density ρ=1800 kg/m3), showing lensing effect as in (A). Color scale ranges from dark blue (vanishing amplitude of displacement field) to red (maximum amplitude). Note the perfectly matched layers that model a plate of infinite transverse dimensions.

Another effect on ground displacement is the frequency dependence of the horizontal to vertical Fourier spectra ratio when the signal passes through the lens (20 m in width, 40 m in length) made of 23 holes (2 m in diameter, 5 m in depth, triangular grid spacing 7.07 x 7.07 m). In this case, the artificial anisotropy (FIG. 7) reduces the horizontal amplification between 5 and 7 Hz [**5**]. For example, this corresponds to the frequency of the first fundamental mode of a "stiff" i.e. low-rise buildings (less than four floors) while a taller building is more flexible.

thus, theoretically, the modification of the structure's response (i.e inertial effect).

In the continuity of these first experiments on structured soils, a promising way to cause a modification on any seismic disturbance is to create a complete dynamic artificial anisotropy by implementing geometrical elements, full or empty, in the soil.

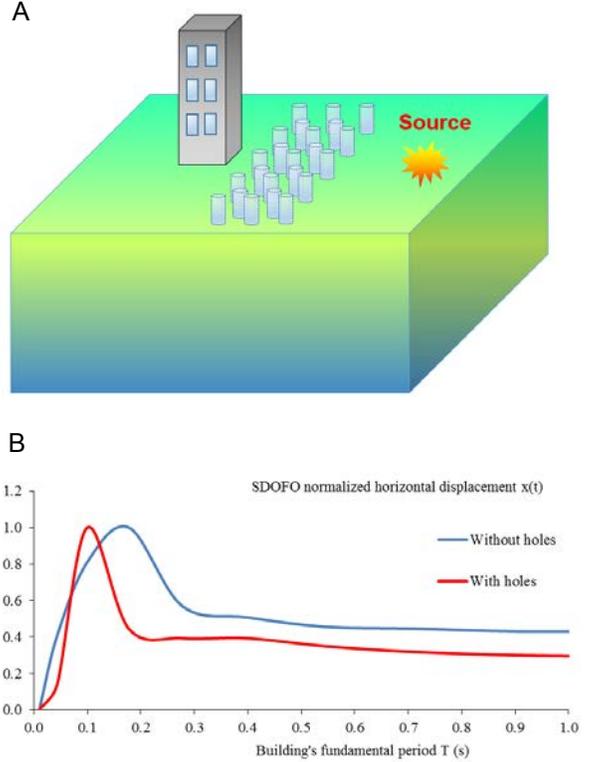

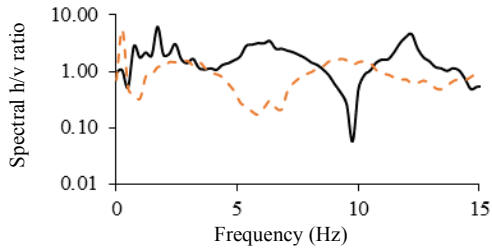

FIG. 7. Experimental test for 23 holes in the soil. Ratio of the horizontal to the vertical components of ground motion versus frequency; seismic land streamer on a soil without holes (black solid line) and same acquisition device for holey soil (orange dotted line). Sensor is located at 50 m from the source point.

The theoretical dynamic response of a building (FIG. 8A) was tested with real seismogram recorded during the 2012 experimental test depicted in FIG. 5. Here, the elastic response spectrum $S(T, \xi)$ is calculated (Eq. 2) for the horizontal displacement $S_d$ of a SDOFO under seismic acceleration of the soil $\ddot{x}_g(t)$. FIG. 8B is the graph of all the relative maximum displacement x(t) of the gravity center of the SDOFO. We vary the value of the structure's period and progressively draw the spectrum for the case without holes (blue line) and the case with holes (red line). The theoretical structure is located at 50 m from de source, just behind the mesh of holes. Results show a bandwidth diminution for the resonance of the structure and a shift towards a lower period value for the maximum horizontal displacement. These results illustrate the significant change of the kinematic effect (§IV) due to the structured soil and

FIG. 8. Experimental test for 23 holes in the soil for theoretical studies of the dynamic response x(t) of a building modeled as a SDOFO (A). Dynamic response calculated in relative horizontal maximum displacement (B). The damping is $\xi_{building} = 0.05$.

The physical process is the interference of waves (body or surface waves) scattered from surfaces or objects. The effects of the dynamic anisotropy are reinforced by the local resonance of implemented, which are disposed along a grid according to transformation elastodynamics ([**40**] and [**43**]) and morphing tools [**42**] that could theoretically lead to an ideal cloak detouring waves around a protected area (FIG. 9). On this occasion, and in order to explore the outstanding peculiarities of crystallography, the researchers also test motifs with five-fold symmetry such as quasi-crystals generated by a cut-and-projection method from periodic structures in higher-dimensional space [**42**].

In this periodic or non-periodic media, the desired effects are total reflection (Bragg's effect), band-gaps, wave-path control, attenuation by energy-dissipation, etc.

In case of vertical columns, numerical simulations have shown zero frequency stop-bands that only exist in the limit of columns of concrete clamped at their base to the bedrock. In a realistic configuration of a sedimentary basin 15m deep we observe a zero frequency stop-band covering a broad frequency range of 0–30 Hz [**41**].

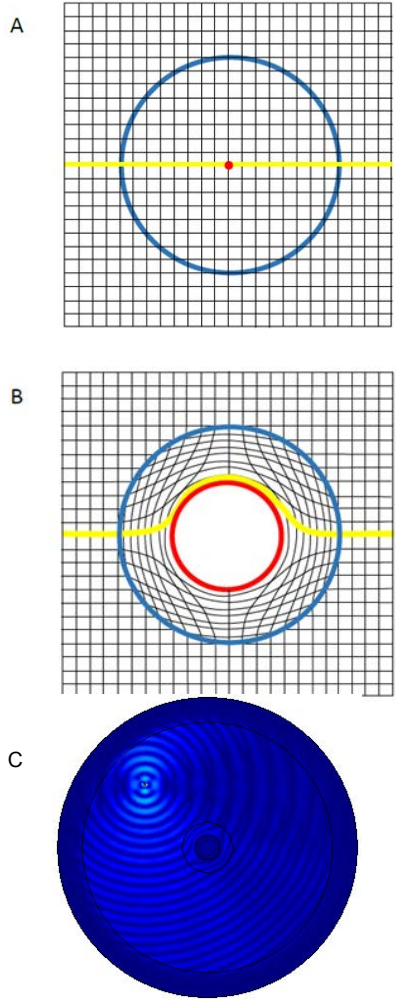

FIG. 9. Illustration of the transformation of a disk into a ring (inspired by [**39**]) from virtual (A) to physical (B) space; (C) Simulation (comsol) of an elastic wave generated by a point force detoured by a seismic cloak. Color scale ranges from dark blue (vanishing amplitude of displacement field) to red (maximum amplitude). Note the perfectly matched concentric layer that models the infinite outer medium.

**Buried Mass-Resonators**

The second group of seismic metamaterial consists of resonators buried in the soil (FIG. 10) in the spirit of tuned-mass dampers (TMD) like those placed atop of skyscrapers. We call this group "Buried Mass-Resonators" (BMR).

We do not consider here the resonators not buried or forming part of the structure. We consider these concepts to be in line with the technologies of seismic isolators.

TMD or BMR is a device consisting of a mass, a spring, and a damper that is attached to a structure for the TMD or in the soil for BMR, in order to respectively reduce the dynamic response of the structure or the soil. The frequency of the damper, or the set of oscillators, is tuned to a particular structural frequency so that when that frequency is excited, the damper will resonate out of phase with the structural motion. Energy is dissipated by the damper inertia force acting on the structure.

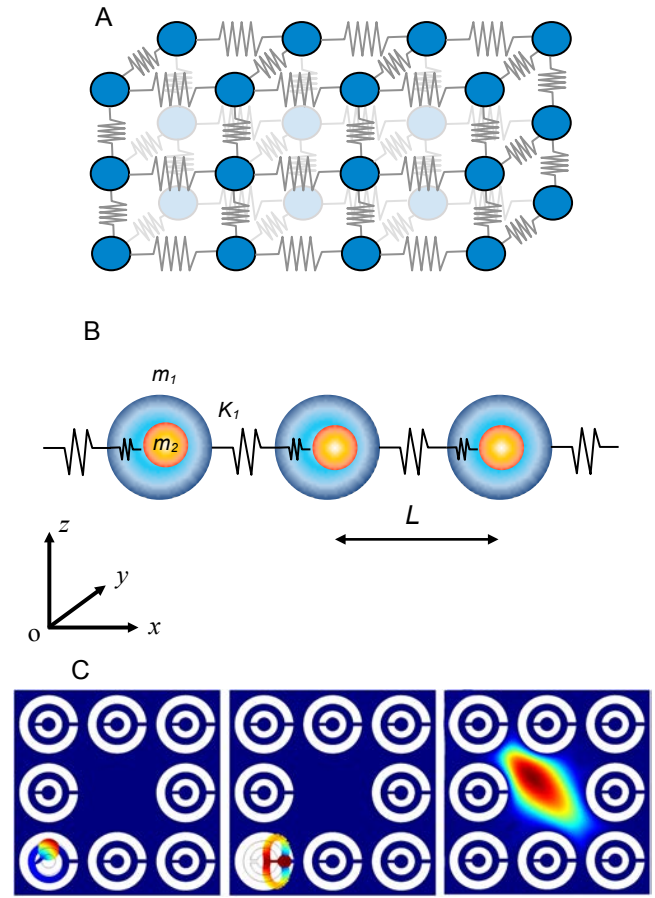

FIG. 10. Top (A), modeling a 3D isotropic elastic solid. Middle (B), modeling the propagation medium consisting of a chain of oscillators with nested masses (1D). Illustration of a single mass $m_1$ with an imbricated mass $m_2$. The set is equivalent to a unique mass $m_{eff}$. The system is under a dynamic loading F(t) in y direction. From [**44**] and [**45**]. Bottom (C), numerical simulations for Floquet-Bloch in-plane elastic waves propagating within a doubly periodic array of supercells (sidelength 6 m) of eight split ring resonators (SRRs) (stress free inclusions in a bulk with soil parameters like in Fig. 6) with a defect, which are the continuous counterpart of discrete models in (A,B). Color scale ranges from dark blue (vanishing amplitude of displacement field) to red (maximum amplitude). This metamaterial creates low frequency stop bands (in the range [1, 10] Hz) associated with localized bending and longitudinal modes within any one of the SRRs, and the defect in the center of the supercell is associated with a defect mode that sits within the stop bands.

Interestingly, authors [46] propose using cross-shaped, hollow and locally resonant (with rubber, steel and concrete), cylinders to attenuate both Rayleigh and bulk waves in the 1–10Hz frequency range. Although this metamaterial seems difficult to implement at a reasonable cost with currently available civil engineering techniques, it certainly opens an interesting route for seismic wave protection.

Unfortunately, the main drawback of this type of locally resonant structure is the difficulty in obtaining very large efficient stop bands; there is always a trade-off between the relative bandwidth and the efficiency of the attenuation, which is directly linked to the quality factor of the resonators. The frequency bandwidth of wave protection can be enlarged by considering arrays of resonant cylinders with different eigenfrequency for two-dimensional stop-bands [47], or cubic arrays of resonant spheres [48] for 3D stop-bands, but such mechanical metamaterials would be hard to implement at the civil engineering scale. Buried isochronous mechanical oscillators have been also envisaged to filter the

**Above-Surface Resonators**

The third type of seismic metamaterial consists of sets of Above-Surface Resonators (ASR).

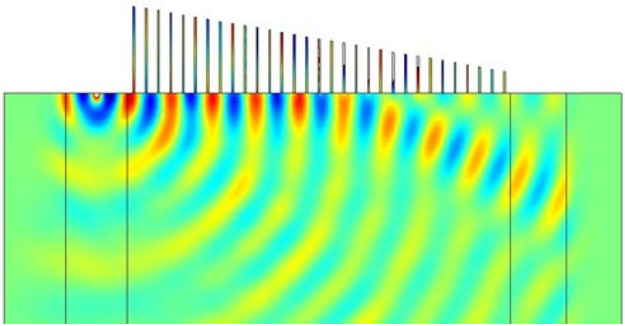

FIG. 11. Simulation of a Rayleigh wave that propagates at a frequency 70 Hz in a forest of trees of decreasing height (14m to 4m) and same diameter (0.3m) atop a soil substrate with elastic parameters like in Fig. 6, which is converted into a downward propagating shear wave by a tree whose longitudinal resonance corresponds to a stop band for the Rayleigh wave at 70 Hz. Note the perfectly matched layers on left and right sides to avoid reflection on computational domain boundary (another layer avoiding reflections at the bottom is not shown). Color scale ranges from dark blue (largest negative value of real part of displacement field) to red (largest positive value of real part of displacement field).

A complementary approach employed in [50, 51] is to draw upon the metamaterials literature that utilizes subwavelength resonators arranged, in their case, upon the surface of an elastic half-space; this results in surface Rayleigh wave to bulk shear wave conversion, see Fig. 11, and surface wave filters with band-gaps but again at higher frequency than those required for seismic protection. Note that if one places the source on the right-hand side of the forest in Fig. 11, the surface wave is simply back-reflected by one of the trees, and no longer converted into a bulk wave. This strongly asymmetric behavior is reminiscent of non-reciprocal media, see [50, 51] for more details. Care needs to be taken with the implementation of such a metawedge as depending upon the incidence of the incoming wave, it might be deleterious for building located on the wrong side of the forest, for instance in a urban environment. Nonetheless, if implemented thoughtfully, this approach might be useful for attenuation of ground vibration caused by human activity such as traffic.

**New dissipative structures**

By another way but with the aim of reducing the effect of the waves on the structure, recent research on gyrobeams implemented in the structure itself [52] open a new perspective in chiral metamaterial design and in a wide range of applications in earthquake wave filtering. It has been demonstrated that the role of gyrobeams is primarily to create low-frequency 'energy sinks', in which waves generated by external excitations are channelled. As a consequence, energy is diverted away from the main structure, which undergoes smaller displacements and smaller stresses.

## VI. METAMATERIAL-LIKE TRANSFORMED URBANISM

In the further development, one can conceive that modern cities with their buildings behave like a group of Above-Surface Resonators [37]. Indeed, when viewed from the sky, the urban fabric pattern appears similar to the geometry of structured devices as metamaterials. Visionary research in the late 1980s [53 – 56] based on the interaction of big cities with seismic signals and more recent studies on seismic metamaterials, has generated interest in exploring the multiple interaction effects of seismic waves in the ground and the local resonances of both buried pillars and buildings.

Techniques from transformational optics and theoretically validate by numerical experiments methods have been used.

The distribution of buildings (made of concrete) within this seismic cloak which is 1km in diameter, with building's height ranging from 10 m to 100 m, which are partially buried in the soil (with a depth of 40 m), mimics the spatially varying refractive index in a conformal invisibility cloak.

## VII. AUXETIC MATERIALS

A promising avenue is study using extreme properties materials. For example, auxetic metamaterials, characterized by their negative Poisson's ratio v and buried in the soil [57] bring bandgaps at frequencies compatible with seismic waves when they are designed appropriately. As a reminder, Poisson's ratio is the ratio of transverse contraction strain to longitudinal extension strain in a stretched bar. Since most common materials become thinner in cross section when stretched, their Poisson's ratio is positive and so negative Poisson ratio materials are unusual. The typical bow-tie element is designed to achieve tunable elastic stop bands for seismic waves.

A remaining track to be explored is the use of these auxetic materials, not only for the foundations but in some parts of the structure itself.

## VIII. CONCLUDING REMARKS

The last decade is marked by the emergence of metamaterials including seismic metamaterials. Well beyond the important scientific and technological advances in the field of structured soils, the concepts of superstructures "dynamically active" under seismic stress and interacting with the supporting soils (soil kinematic effect) and its neighbors is clearly highlighted.

Tools for exploring the design of structured soil configurations have been developed ([41], [42]) and first full-scale experimental tests have shown the influence of these buried structures on the propagation of surface waves ([4], [5]). Three types of seismic metamaterials are distinguished: "Seismic Soil-Metamaterials" (SSM), "Buried Mass-Resonators" (BMR) and Above-Surface Resonators (ASR). At this stage, such mechanical metamaterials as BMR would be hard to implement at the civil engineering scale, except for frequency vibration (>30 Hz) generated by human activities. There are good prospects for SSM because the process is capable of producing large quantities of elements in the soil, and ASR, especially if their effects can be coupled.

In parallel, new possibilities to dissipate the seismic energy in the superstructures themselves are emerging thanks to the contributions of auxetic materials [57] and metamaterials [52].

## ACKNOWLEDGEMENTS

S.G. wishes to thank ERC funding (ANAMORPHISM project) during 2011-2016 that helped strengthen collaboration between the authors and current ANR funding (METAFORET project).